\documentclass[12pt,a4paper]{article}
\usepackage{amsmath}
\usepackage[dvips]{graphicx}
\newcommand{\bc}{\begin{center}}
\newcommand{\ec}{\end{center}}
\newcommand{\be}{\begin{equation}}
\newcommand{\ee}{\end{equation}} 
\newcommand{\ba}{\begin{eqnarray}}
\newcommand{\ea}{\end{eqnarray}}
 
\begin{document}
\begin{flushright} INFN-GEF-TH-2/2013 \end{flushright}
 \baselineskip 24pt
 \bc {\Large \bf On the gravitational self-energy of a spherical shell}

G. Dillon\footnote{e-mail: dillon@ge.infn.it} \\ 
\baselineskip 16pt
{\it Dipartimento di Fisica, Universit\`a di Genova\\
INFN, Sezione di Genova} \ec

 \baselineskip 14pt
 \noindent{Abstract: According to Einstein's  mass-energy equivalence, a body with a %%@
given mass extending in a large region of space, will get a smaller mass when confined %%@
into a smaller region, because of its own gravitational energy. The classical %%@
self-energy problem has been studied in the past in connection with the renormalization %%@
of a charged point particle. Still exact consistent solutions  have not been  %%@
thoroughly discussed in the simpler framework  of  Newtonian gravity. Here we  exploit %%@
a spherical symmetrical shell model and find two possible solutions, depending on some %%@
additional assumption. The first solution goes back to Arnowitt, Deser and Misner %%@
(1960). The second  is new and yields a new vanishingly  small value ($10^{-55}cm$) for %%@
the classical electron radius.}  
 \vskip10pt

\baselineskip 15pt 
\section{Introduction}
 In classical Newtonian physics, the gravitational energy of a simple  spherically %%@
symmetric shell of mass $M_0$ and radius $R$ is
\be U_0(R)=-G\frac{M_0^2}{2R}
\label{ess}\ee
where $G$ is the gravitational constant.

Taking into account mass-energy equivalence from special relativity, this binding %%@
energy (negative) is equivalent to a mass defect. Furthermore the equality of inertial %%@
and gravitational masses has been tested experimentally even in presence of sizeable %%@
mass defects due to large  binding energies \cite{Sau,Zee}. Hence the mass of the shell %%@
will be different from $M_0$ and, in turn, (\ref{ess}) should be reexpressed in terms %%@
of the new ``renormalized" mass. In the following we shall refer to $M_0$ as the %%@
``bare" mass while the renormalized $M(R)$ is the resulting mass when $M_0$ is %%@
uniformly distributed on a spherical shell of radius $R$. $M(R)$ takes into account the %%@
gravitational self-energy, while $M_0$ corresponds to the sum of  all the masses that %%@
one would obtain tearing the sphere in many small pieces and moving them  away %%@
apart\footnote{Here we are disregarding any form of kinetic energy or internal pressure %%@
or stress that can also contribute to the mass.}. 

The renormalized mass at the first order in $c^{-2}$ is immediately obtained from %%@
(\ref{ess})
\be M(R)\approx M_0+U_0(R)/c^{-2}= M_0(1-G\frac{M_0}{2Rc^2})\label{1o}\ee
However  the exact calculation of  $M(R)$ from
a given bare mass $M_0$ is not trivial and rests upon some additional assumption.

The classical self-energy problem has been studied in the past by R. Arnowitt, S. Deser %%@
and C.W. Misner (ADM) \cite{ADMprl, ADMiv, ADM} in the framework of a canonical %%@
formulation of General Relativity. Here we whish to consider the problem in the simpler %%@
context of Newtonian physics with minimal assumptions. In this sense the present paper %%@
has some pedagogical character. In Sec.2 we revisit the solution given by ADM. We find %%@
however that this solution is inconsistent with the following expression
\be \delta U(R)=-G\frac{M(R)\delta m_0}{R}
\label{2}\ee  
for the gravitational interaction energy between the shell and a test-particle $\delta %%@
m_0$ settled on its surface.
  In Sec.3 we relate the reason for the inconsistency to the lack of an appropriate %%@
mass renormalization of the test-particle due to $\delta U(R)$ itself. In Sec.4 we %%@
propose an alternative solution, consistent with (\ref{2}), which allows also for a new %%@
definition of the classical electron radius, which adds to the well known one.
%%%%%%%%%%%%%%%%%%%%%%%%%%%%%%%%%%%%%%%%%%%%%%%%%%%%

\section{A consistent solution for the spherical shell and an inconsistency}
Given the expression (\ref{ess}) for the gravitational energy of a spherical shell, it %%@
seems quite natural to write down the following consistent equation for the %%@
renormalized mass $M(R)$ 
\be M(R)=M_0-\frac{G}{2}\frac{M(R)^2}{Rc^2}
\label{ADM}\ee
or equivalently for the gravitational self-energy of the shell
\be U(R)=-G\frac{(M_0+U(R)/c^2)^2}{2R}\label{eqU}\ee
Defining 
\be R_0=\frac{GM_0}{c^{2}}\label{R0}\ee
one gets the (positive) solution for the mass
\be M(R)=M_0(-1+\sqrt{1+2R_0/R}\ )R/R_0
\label{sol1}\ee
with the corresponding gravitational self-energy
\be U(R)=-G\frac{M(R)^2}{2R}=-M_0c^2(1+R_0/R-\sqrt{1+2R_0/R}\ )R/R_0 \label{solU1}\ee
The equation (\ref{ADM}) has been considered \cite{ADM, M} in the framework of the %%@
classical theory of the electron. In fact, adding to (\ref{ADM}) the contribution to %%@
the mass of the electromagnetic energy $e^2/2R$ (this time positive), the ensuing %%@
solution tends to a finite value  when $R\rightarrow 0$:
 \be M(R\rightarrow 0)=|e|/\sqrt{G}\equiv m_G\label{mG}\ee
 independent of $M_0$. This elegant result exhibits a nice feature of the gravitational %%@
self-energy as a natural cutoff for the Coulomb self-energy of a point charge. The %%@
result is however numerically too big ($m_G\approx 10^{21}m_e$) compared to the %%@
electron mass. We shall  further comment on this.

Now one could naively think that  the (now renormalized) shell yields a gra\-vitational %%@
field identical to that of a spherical shell with the trivial substitution %%@
$M_0\rightarrow M(R)$, i.e. a gravitational potential function (for $r\ge R$)
\be \Phi(r)=-G\frac{M(R)}{r}
\label{1}\ee 
Hence the interaction energy with a test particle settled on its surface should be %%@
given by (\ref{2}). If so, the total mass of the system (spherical shell of %%@
renormalized mass $M(R)$ plus test-particle $\delta m_0$ on its surface) is
\be M_t(R)=M(R)+\delta m_0+\delta U(R)/c^2=M(R)+\delta %%@
m_0(1-G\frac{M(R)}{c^2R})\label{Mt}\ee
Substituting $M(R)$ from (\ref{sol1}) one has
\be M_t(R)= M(R)+\delta m_0(2-\sqrt{1+2R_0/R}\ )\label{Mt1}\ee

Here we come to a contradiction. Indeed, suppose we want to deposit a test particle %%@
$\delta m_0$ on the surface of $M(R)$ and let us think about this test mass as being %%@
uniformly distributed on a thin spherical shell of radius $r$ centered on the origin, %%@
just as $M(R)$. (Note that, neglecting higher orders in $\delta m_0$, we do not worry %%@
about self-energy of $\delta m_0$ on its own. In other words: $\delta m(r)\approx %%@
\delta m_0$.) Now imagine to bring $r$ to $R$ and to stick $\delta m_0$ as a thin film %%@
on   $M(R)$. Then, viewing the system as a new shell of bare mass $M_0+\delta m_0$, one %%@
has again from (\ref{sol1})
\be M_t(R)=M(R)+\frac{\partial M(R)}{\partial M_0}\delta m_0= M(R) +\frac{\delta %%@
m_0}{\sqrt{1+2R_0/R}}
\label{Mt1'}\ee
that does not agree with (\ref{Mt1}), but at the first order in $R_0/R$ (i.e. in %%@
$c^{-2}$). So there is a mistake somewhere.

%%%%%%%%%%%%%%%%%%%%%%%%%%%%%%%%%%
 
\section{Renormalizing the test-mass}
 In Sec.2 we thought about the test-particle as being spherically distributed on a thin %%@
shell concentric with the shell $M(R)$. We argued that we should not worry about its %%@
own self-energy (since we work at first order in $\delta m_0$); however we left aside %%@
the possibility of further renormalization consequences on the masses due to the %%@
interaction energy $\delta U(R)$ itself. Let us assume, by now, that the whole  $\delta %%@
U(R)$ be attributed to  the test-mass, i.e.
\be \delta m_0\rightarrow \delta m=\delta m_0+\delta U(R)/c^2\label{rindm}\ee 
This assumption amounts rewriting  (\ref{2}) as a self-consistent equation for $\delta %%@
U(R)$
\be \delta U(R)=-G\frac{M(R)(\delta m_0+\delta U(R)/c^2)}{R}
\label{3}\ee  
From (\ref{3})
\be \delta U(R)=-G\frac{M(R)\delta m_0}{R(1+G\frac{M(R)}{Rc^2})} 
\label{dU2}\ee
and 
\be M_t(R)=M(R)+\frac{\delta m_0}{1+G\frac{M(R)}{Rc^2}}\label{Mt2}\ee
at variance with (\ref{Mt}). Substituting (\ref{sol1}) we get
\be 1+G\frac{M(R)}{Rc^2}=\sqrt{1+2R_0/R}\ee
so (\ref{Mt2}) now agrees with (\ref{Mt1'}) and there is no inconsistency.

In fact  the starting equation (\ref{ADM})  may be obtained from (\ref{Mt2}) as %%@
follows: Viewing the total system as a new shell with increased bare mass ($M_0+\delta %%@
m_0$), one can write $$M_t(R)-M(R)\equiv dM(R) \quad ;\quad \delta m_0\equiv dM_0$$ 
  Hence (\ref{Mt2}) is a differential equation that yields the mass $M(R)$ of a %%@
spherical shell of radius $R$ as a function of its bare mass $M_0$:
\be \frac{dM(R)}{dM_0}=\frac{1}{1+G\frac{M(R)}{Rc^2}}
\label{E2}\ee
whose solution is just (\ref{ADM}).
 
Note that the renormalization of  $\delta m_0$ may be equivalently described in terms %%@
of a suitable modification of the gravitational potential, that, instead of (\ref{1}), %%@
should be written as (for $r>R$)
\be \Phi(r)\rightarrow\tilde\Phi(r)=-\frac{G}{(1+G\frac{M(R)}{rc^2})}\frac{M(R)}{r}
\label{rgp}\ee
with $M(R)$ given by (\ref{sol1}).
%%%%%%%%%%%%%%%%%%%%%%%%%%%%%%%%%%%
\section{An alternative solution and a new classical electron radius}
In Sec.3 it was proved that renormalizing the mass of the test-particle by the %%@
interaction energy with the massive shell leads to (\ref{sol1},\ref{solU1}). However %%@
this assumption may appear somewhat arbitrary. Moreover it seems not conciliable with %%@
additivity. In fact let us think about the shell $M(R)$ as made up by the sum of a %%@
large number $N$ of light overlapping shells each of mass $\delta m'$ 
$$ \sum_i\delta m_i'=N\delta m'\equiv M(R)$$
and rewrite (\ref{2}) as the sum of the interaction energies with each sub-shells
\be \delta U(R)=-G\sum_i\frac{\delta m_i'\delta m_0}{R}\equiv \sum_i \delta %%@
u_i(R)=N\delta u(R)
\label{sum}\ee
Now for each individual term in the sum in (\ref{sum}) let us perform the %%@
renormalization of the test mass as before ($\delta m_0\rightarrow \delta m_0 + \delta %%@
u(R)/c^2$). Then
 \be \delta U(R)=-G\sum_i\frac{\delta m_i'(\delta m_0+\delta u(R)/c^2)}{R}\approx %%@
-G\frac{M(R)\delta m_0}{R}
\label{sum'}\ee
for large N. Moreover the final result (\ref{sum'}) will not change even if the little %%@
mass equivalent to the interaction energy were attributed  (fully or in part) to the %%@
shell $M(R)$ rather than to the test-particle. This is because the ensuing correction %%@
to $\delta U(R)$ comes to be at a higher order in $\delta m_0$ in this case. 

Then, assuming additivity, one should not change (\ref{2}) and look for a solution %%@
other than (\ref{sol1},\ref{solU1}). In fact we may exploit the method outlined in %%@
Sec.3. Starting from (\ref{Mt}) we get the differential equation
\be \frac{dM(R)}{dM_0}=1- G\frac{M(R)}{Rc^2}
\label{E1}\ee
whose solution is
\be \ln\big(1-G\frac{M(R)}{Rc^2}\big)=-G\frac{M_0}{Rc^2}
\label{e1}\ee
\be M(R)=\frac{Rc^2}{G}\big(1-\exp[-G\frac{M_0}{Rc^2}]\big)\equiv %%@
M_0\big(1-\exp[-\frac{R_0}{R}]\big)\frac{R}{R_0}
\label{sol1'}\ee
Instead of (\ref{Mt1}) we have now 
\be M_t(R)=M(R)+\delta m_0\exp[-\frac{R_0}{R}]    \label{Mt1''}\ee
and there is, of course, no contradiction.

The alternative solution (\ref{sol1'}) deserves a couple of comments in connection with %%@
the classical theory of the electron \cite{J}. To obtain the mass of a spherical shell %%@
of radius $R$, charge $e$ and bare mass $M_0=0$, it is enough to make the substitution  %%@
$M_0\rightarrow e^2/2Rc^2$ in (\ref{sol1'}). One gets
\be M_e(R)=\frac{c^2R}{G}\big(1-\exp[-G\frac{e^2}{2R^2c^4}]\big) \label{me1}  \ee
Again we may appreciate the regularizing role of the gravitational self-energy that %%@
heals the divergency of the Coulomb self-energy when $R\rightarrow 0$. Instead of the %%@
previously quoted result $M_e(R\rightarrow 0)=m_G=|e|/\sqrt{G}$ \cite{ADM, M}, this %%@
time we obtain $M_e(R\rightarrow 0)=0$. 

Let us now define
\be R_e\equiv Gm_e/c^2 \label{Re}\ee
and rewrite (\ref{me1}) in terms of $R_e$ and $r_e=e^2/m_ec^2$ (the usual expression %%@
for the classical electron radius \cite{J}). We get
\be M_e(R)=m_e\frac{R}{R_e}\big(1-\exp[-\frac{r_eR_e}{2R^2}]\big) \label{me2}  \ee
In order that $M_e(R)\equiv m_e$, one has to find solutions of  
\be \frac{R}{R_e}\big(1-\exp[-\frac{r_eR_e}{2R^2}]\big)=1 \label{eqR}  \ee
Numerically for the electron it is: $r_e=2.82\cdot 10^{-13}cm$ and $R_e=0.7\cdot %%@
10^{-55}cm$. Then it is immediately seen that (\ref{eqR}) has two distinct solutions %%@
($R_1,R_2$), almost exactly coincident with $r_e/2$ and $R_e$. Note that, while for the %%@
first solution ($R_1\approx r_e/2$)  gravitational self-energy effects are totally %%@
negligible, the opposite happens for $R_2\approx R_e$ (see (\ref{Re})). 
 
%%%%%%%%%%%%%%%%%%%%%%%%%%%%%%%%%%%%%%%%%%%%%%%%
\section{Concluding remarks}
In this paper, taking due account of mass-energy equivalence (and of the equality of %%@
the inertial and gravitational masses), we faced the seemingly trivial problem of %%@
calculating the gravitational self-energy of a given distribution of mass. We exploited %%@
the simple model of a spherically symmetrical distribution over a shell of radius $R$. %%@
The problem may equivalently be stated as how to calculate the renormalized mass of the %%@
given distribution $M(R)$ from its bare mass $M_0$. We displayed two possible %%@
solutions. The first one was known since 1960 \cite{ADM, M}, the second is new. The %%@
difference between the two is related to specific assumptions about mass %%@
renormalization. The first one requires a modification (see (\ref{rgp})) of the %%@
gravitational potential generated by the shell with respect to the well known %%@
expression of Newtonian theory (\ref{1}). The second solution does not need any %%@
modification of (\ref{1}) and follows from a simple assumption of additivity (see %%@
Sec.4). 

When the bare mass $M_0$ is substituted with $e^2/2Rc^2$ one gets the gravitationally %%@
renormalized e.m. mass of a pure static electric charge uniformly distributed on a %%@
spherical shell of radius $R$. Defining
\be R_G = |e|\sqrt{G}/c^2\equiv Gm_G/c^2 \label{RG}\ee
 it is for the solution (\ref{sol1})
 \be M_e(R)=m_G\frac{R}{R_G}\big(-1+\sqrt{1+(R_G/R)^2}\big)\label{me3}\ee
 which reaches its maximum value $m_G=|e|/\sqrt{G}$ at $R=0$. On the other hand from %%@
the alternative solution (\ref{me1}) one gets
\be  M_e(R)=m_G\frac{R}{R_G}\big(1-\exp[- \frac{R_G^2}{2R^2}]\big) \label{me4}  \ee
that has a maximum of the order of $m_G$ at $R\approx R_G$. Since (\ref{me4}) goes to %%@
zero when $R\rightarrow 0$, one finds a further solution for the electron mass at a %%@
vanishingly small radius ($10^{-55}cm$).

Indeed in \cite{M} it was stated that a system cannot have pointlike dimensions at the %%@
mass $m_G$ and the minimum possible dimensions for that value are just given by %%@
(\ref{RG}). It is remarkable that $m_G$ may also be written as $m_G=\sqrt{\alpha}\ %%@
M_{Planck}$; likewise $R_G=\sqrt{\alpha}\ l_{Planck}$. As emphasized in \cite{Vi}, %%@
these quantities are purely classical (the square root of the fine-structure constant %%@
$\alpha$ cancels $\hbar$) and define the scale at which gravitational and %%@
electromagnetic self-energies become comparable.
%%%%%%%%%%%%%%%%%%%%%%%%%%%%%%%%%%%%%%%%%%%%%%%%%
 %%%%%%%%%%%%%%%%%%%%%%%%%%%%%%%%%%%%%%%%%%%%%
 
%%%%%%%%%%%%%%%%%
\end{document}